\documentclass{article}
\usepackage[final]{graphics}
\usepackage{bbm}

\font\tenmsa=msam10
\font\sevenmsa=msam7
\font\fivemsa=msam5
\font\tenmsb=msbm10
\font\sevenmsb=msbm7
\font\fivemsb=msbm5
\newfam\msafam
\newfam\msbfam
\textfont\msafam=\tenmsa  \scriptfont\msafam=\sevenmsa
\scriptscriptfont\msafam=\fivemsa
\textfont\msbfam=\tenmsb  \scriptfont\msbfam=\sevenmsb
\scriptscriptfont\msbfam=\fivemsb

\global\mathchardef\lesssim "142E

\newcommand{\slL}{\raise.15ex\hbox{$/$}\kern-.53em\hbox{$L$}}
\newcommand{\slP}{\raise.15ex\hbox{$/$}\kern-.53em\hbox{$P$}}
\newcommand{\slR}{\raise.15ex\hbox{$/$}\kern-.53em\hbox{$R$}}
\newcommand{\slQ}{\raise.15ex\hbox{$/$}\kern-.53em\hbox{$Q$}}
\newcommand{\slK}{\raise.15ex\hbox{$/$}\kern-.53em\hbox{$K$}}
\newcommand{\slSigma}{\raise.15ex\hbox{$/$}\kern-.53em\hbox{$\Sigma$}}
\newcommand{\slcalP}{\raise.15ex\hbox{$/$}\kern-.63em\hbox{$\cal P$}}

\newcommand{\be}{\begin{equation}}
\newcommand{\ee}{\end{equation}}     
\newcommand{\bea}{\begin{eqnarray}}
\newcommand{\ena}{\end{eqnarray}}

\def\build#1\over#2{\mathrel{\mathop{\kern 0pt#1}\limits_{#2}}}

\font\tenimbf=cmmib10 at 12pt
\font\sevenimbf=cmmib10 at 7pt
\font\fiveimbf=cmmib10 at 5pt
\newfam\imbf
\textfont\imbf=\tenimbf
\scriptfont\imbf=\sevenimbf
\scriptscriptfont\imbf=\fiveimbf
\def\imb{\fam\imbf\tenimbf}

\begin{document}
\begin{titlepage}
\title{
\bf{Landau-Pomeranchuk-Migdal effect\\
in thermal field theory}}

\author{
P.~Aurenche$^{(1)}$, F.~Gelis$^{(2)}$, H.~Zaraket$^{(1)}$}
\maketitle

\begin{center}
\begin{enumerate}
\item Laboratoire de Physique Th\'eorique LAPTH,\\
UMR 5108 du CNRS, associ\'ee \`a l'Universit\'e de Savoie,\\
BP110, F-74941, Annecy le Vieux Cedex, France
\item Brookhaven National Laboratory,\\
Physics Department, Nuclear Theory,\\
Upton, NY-11973, USA
\end{enumerate}
\end{center}

\begin{abstract}
  In recent studies, the production rate of photons or lepton pairs by
  a quark gluon plasma has been found to be enhanced due to collinear
  singularities. This enhancement pattern is very dependent on rather
  strict collinearity conditions between the photon and the quark
  momenta. It was estimated by neglecting the collisional width of
  quasi-particles. In this paper, we study the modifications of this
  collinear enhancement when we take into account the possibility for
  the quarks to have a finite mean free path. Assuming a mean free
  path of order $(g^2T\ln(1/g))^{-1}$, we find that only low invariant
  mass photons are affected. The region where collision effects are
  important can be interpreted as the region where the
  Landau-Pomeranchuk-Migdal effect plays a role in thermal photon
  production by bremsstrahlung.  It is found that this effect modifies
  the spectrum of very energetic photons as well.  Based on these
  results and on a previous work on infrared singularities, we end
  this paper by a reasonable physical picture for photon production by
  a quark gluon plasma, that should be useful to set directions for
  future technical developments.

\end{abstract}
   \vskip 4mm
\centerline{\hfill LAPTH--790/2000, BNL-NT--00/11}

\vfill
\thispagestyle{empty}
\end{titlepage}

\section{Introduction}
The photon production rate is thought to be a quantity of
phenomenological interest in heavy ions collisions, possibly enabling
one to detect the formation of a quark-gluon plasma. Part of the
interest in this electromagnetic observable comes from the fact that
photons are relatively weakly coupled to nuclear matter
($\alpha_{_{EM}}\ll \alpha_{_{S}}$). Given the typical size of the
system in such collisions (much smaller than the mean free path of a
photon), they do not re-interact between their production and their
observation. As a consequence, photons (real photons, or virtual
photons decaying eventually into a lepton pair) can provide
information on the state of the system at the time they were produced.

In order to calculate the photon yield from a hot quark-gluon plasma,
thermal field theory is the tool of choice since its Feynman rules
automatically take into account the presence of a thermal bath with
the appropriate distributions of partons.  Thermal gauge theories are
however plagued by infrared singularities arising from the
Bose-Einstein distribution functions which are singular at zero
energy.  An improvement over the bare Feynman rules is achieved by the
resummation of one-loop thermal contributions known as hard thermal
loops (HTL) \cite{Pisar2,BraatP1,BraatP2,FrenkT1,FrenkT2}.  These
thermal corrections make it possible to include in the propagators
effects like Debye screening or Landau damping, and transform partons
into massive quasi-particles. From a quantitative perspective, they
provide important changes to the dynamics of soft modes (of momentum
of order $gT$ or less). This resummation lacks however two features:
it does not provide any Debye screening for static magnetic fields
(such a screening is expected to arise non-perturbatively in QCD at
the length scale $(g^2T)^{-1}$), and its quasi-particles do not
undergo collisions (their collisional\footnote{In a plasma, it is very
  important to distinguish two contributions to the total width: the
  {\sl decay width} made of the zero temperature contribution to the
  imaginary part of the quark self-energy, and the {\sl collisional
    width} which exists only in a medium. The latter is also called
  ``anomalous damping rate'' (or just ``damping rate'' for short) and
  is of order $g^2T\ln(1/g)$, while in QED/QCD the former starts at
  the order $g^4$. The width $\Gamma$ that we introduce in this paper
  is the collisional width, and its inverse is the mean free path of
  the quark in the plasma.} width is also a sub-leading effect of
order $g^2T\ln(1/g)$).

In the framework of thermal field theory, the photon/dilepton rate is
obtained via the calculation of the imaginary part of the retarded
photon polarization tensor \cite{Weldo3,GaleK1}. This object has been
evaluated at one-loop in the HTL-improved perturbative expansion, both
for virtual \cite{BraatPY1,Wong1} and real photons
\cite{KapusLS1,BaierNNR1,AurenBP1,BaierPS1,Niega6}.  More recently,
new processes like bremsstrahlung were studied in detail in this
framework and have been found to be dominant sources of low invariant
mass photons \cite{AurenGKP1,AurenGKP2,AurenGKZ1}. Despite the fact
that this process arrives only in 2-loop contributions to the photon
polarization tensor, it is always important because of a strong
collinear enhancement.  Indeed, it was found in \cite{AurenGKP2} that
for a very small photon invariant mass, the corresponding diagram
contains collinear singularities that, when regularized by a thermal
quark mass of order $gT$, give an enhancement by a factor of order
$1/g^2$ over naive estimates coming from power counting.

After finding that certain 2-loop terms are contributing at leading
order, one may wonder whether this result is specific to this 2-loop
contribution only or if, on the contrary, this is an indication of the
breakdown of perturbative expansion (even improved with HTLs).  In a
recent paper \cite{AurenGZ1}, we studied what is the effect of loop
corrections to this 2-loop diagram. Power counting indeed indicates a
problem very similar to the problem raised by Linde \cite{Linde1} for
the calculation of the free energy, due to the lack of Debye screening
for static magnetic modes. This problem is avoided for the production
of massive enough photons, because some cancellations (occuring within
any given topology, when one is summing over all the cuts contributing
to the imaginary part) generate a kinematical cutoff large enough to
prevent any sensitivity to the non-perturbative scale $g^2T$.
Unfortunately, this cutoff is smaller than $g^2T$ whenever the
invariant mass of the produced photon is too small (typically
$Q^2<g^2Tq_0$ for $q_0<T$). In this low invariant mass region, the
photon rate is therefore non perturbative: exchanged transverse gluons
reach the scale $g^2T$ of the non-perturbative ``magnetic mass'', and
an infinite set of diagrams must be resummed.

In this paper, we present a completely different approach to higher
order corrections, that completes the picture outlined in
\cite{AurenGZ1}.  The idea behind the present study is that a width on
the quark propagator will act as a regulator in the collinear sector,
because it moves the poles of the propagator away from the real energy
axis. Such a collisional width is necessarily a higher-loop effect,
because the hard thermal loop framework does not take into account the
collisions of quasi-particles.  Having in mind the fact that 2-loop
contributions are important because of collinear enhancement, an
important question to answer is how much of this enhancement is lost
when an additional regulator like a width is taken into account. This
is the question we want to address in this paper, by calculating the
same 2-loop diagrams as in \cite{AurenGKP1,AurenGKP2}, in the presence
of a quark width. To be more definite, and keep the model as well as
the calculations simple, we use a momentum-independent width.

A word of caution is necessary here: the formulae found using this
simple model should not be taken as an accurate quantitative account
of what the effect of such a width will be on thermal photon
production rates.  Indeed, the constant width model is probably too
naive to be realistic, and more importantly our calculation disregards
the fact that a modification of the vertices should in principle
accompany the modification of the quark propagators.  Nevertheless,
this simple approach is sufficient here for our purpose which is just
to determine the region in which effects of a width of order
$g^2T\ln(1/g)$ are to be expected, since that gives another handle on
how and when higher order corrections may affect photon production by
a quark-gluon plasma.

We find that the region where a width of order $g^2T\ln(1/g)$ is
important is very similar to the one found in \cite{AurenGZ1} for the
contribution of higher-loop topologies due to an IR sensitivity to the
scale of the magnetic mass. Despite their similarity, these two non
perturbative regions have different physical interpretations. In
particular, we find that the sensitivity to the collisional width of the quarks is
a manifestation of the Landau-Pomeranchuk-Migdal effect. Indeed, it
occurs when the formation time of the photon is larger than the
mean free path of the quark producing the photon. An interesting consequence
of our study is that the LPM effect also modifies the spectrum of
highly energetic photons.

The structure of the paper is as follows. Section \ref{sec:model}
makes more precise our modeling of the mean free path for the
quarks. In section \ref{sec:1-loop}, we start by computing the 1-loop
contribution in presence of a width.  Although not related to
collinear singularities in any way, the purpose of this warmup
exercise is two-fold: it illustrates the technology (how one does
calculations with a width), and it shows in a simple way how
collisions can open up the phase space. We also show that this 1-loop
contribution is canceled by the resummation of vertex corrections, and
is therefore not physical.

In section \ref{sec:2-loop-calcul}, we present the 2-loop calculation
with a quark width, and obtain a rather simple generalization of the
formulae of \cite{AurenGKP2}. These formulae are discussed extensively
in section \ref{sec:2-loop-discussion}, in which we also determine the
region where a width of order $g^2T\ln(1/g)$ plays an important role.

Section \ref{sec:LPM} is devoted to establishing the connection
between the previous results and the LPM effect.  In section
\ref{sec:annihilation}, we study the process obtained from
bremsstrahlung by crossing symmetry, which turns out to be important
in the region of large photon energy. We show that this process is
also affected by the LPM effect.  The last section contains concluding
remarks. In particular, we combine the present work on the LPM effect
with previous results on infrared singularities \cite{AurenGZ1} in
order to make a syntesis and present a reasonable physical picture of
thermal photon production.

\section{Model}
\label{sec:model}
Let us first make our framework more definite. The only modification
compared to the context extensively described in \cite{AurenGKP2} is
that the quark propagators are given a width, as outlined by the
following substitution in the retarded and advanced propagators:
\begin{equation}
\Delta_{_{R,A}}(P)\equiv{1\over{P^2-M^2_\infty\pm ip_0\epsilon}}
\to{\imb\Delta}_{_{R,A}}(P)\equiv{1\over{(p_0\pm 
i\Gamma)^2-{\imb p}^2-M^2_\infty}}\; ,
\label{eq:propagator}
\end{equation}
where $M_\infty\sim gT$ is the usual asymptotic thermal mass for hard
quarks \cite{FlechR1,AurenGKP2}, and where $\Gamma$ is a constant
width. The expression in Eq.~(\ref{eq:propagator}) is sufficient for
the physics we consider, which is dominated by hard quark momenta
close to the mass shell.  Whenever we need to estimate the order of
magnitude of a term containing $\Gamma$ in the following, we assume
that it is of order $g^2T\ln(1/g)$ (with $g\ll1$).  Note that we could
as well have written
\begin{equation}
{\imb \Delta}_{_{R,A}}(P)={1\over{P^2-M^2_\infty\pm2ip_0\Gamma}}
\end{equation}
since the two differ by the small correction $\Gamma^2\ll M^2_\infty$
to the real part of the denominator.  The retarded (resp.  advanced)
propagator has two complex poles in the $p_0$ plane, located at
$p_0=\pm\omega_{\imb p}-i\Gamma$ (resp.  $p_0=\pm\omega_{\imb
  p}+i\Gamma$), with $\omega_{\imb p}\equiv\surd({\imb
  p}^2+M^2_\infty)$.

\section{1-loop study}
\label{sec:1-loop}
\subsection{Calculation}
\begin{figure}[htbp]
\centerline{\resizebox*{!}{3.75cm}{\includegraphics{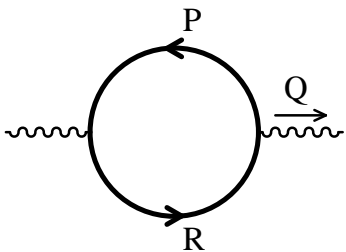}}}
\caption{\sl 1-loop diagram. The boldface 
  quark line denotes the resummation of a width on the fermion
  propagator.}
\label{fig:1-loop}
\end{figure}

Let us now consider the simplest calculation conceivable in this
framework, namely the computation of the 1-loop contribution to the
photon polarization tensor. Our purpose is simply to illustrate
possible changes brought in when taking into account a width for the
quarks running in the loop. We do not attempt a complete calculation
including HTL vertices, as pictured on Fig.~\ref{fig:1-loop}.
The contribution of this diagram to the imaginary part of the retarded
polarization tensor is trivial to obtain:
\begin{eqnarray}
&&\hbox{\rm Im}\,\Pi^\mu{}_\mu(Q)\approx{1\over 2}
e^2 \int {{d^4 P}\over{(2\pi)^4}} 
\left[n_{_{F}}(r_0)-n_{_{F}}(p_0)\right]\;
\hbox{\rm Tr}\nonumber\\
&&\qquad\qquad\qquad \qquad\times[{\imb\Delta}_{_{R}}(P)
-{\imb\Delta}_{_{A}}(P)]
[{\imb\Delta}_{_{R}}(R)
-{\imb\Delta}_{_{A}}(R)]\ ,
\label{eq:1-loop}
\end{eqnarray}
where ${\rm Tr}$ denotes the Dirac's trace associated to the quark loop:
\begin{equation}
{\rm Tr}=-8 P\cdot R\; .
\end{equation}

The case with $\Gamma=0$ is made simple by the fact that the
differences ${\imb \Delta}_{_{R}}-{\imb \Delta}_{_{A}}$ are $\delta$
functions, thereby enabling some of the integrals to be performed
trivially. Let us remind that for the case $Q^2 < 4M^2_\infty$ which
we assume throughout this paper, ${\rm Im}\,\Pi{}^\mu{}_\mu(Q)=0$ if
$\Gamma=0$ due to incompatibilities between the $\delta$ functions.
To evaluate Eq.~(\ref{eq:1-loop}) when $\Gamma\not= 0$, we perform the
integral over $p_0$ by closing the real axis in the complex energy
plane and using the theorem of residues.

Note first that one can disregard the poles of the statistical weights
$n_{_{F}}(p_0)-n_{_{F}}(r_0)$. Indeed, these poles are the imaginary
fermionic Matsubara frequencies and are to be plugged into differences
like ${\imb \Delta}_{_{R}}(P)-{\imb \Delta}_{_{A}}(P)$. The fact that
$\Gamma\ll T$ makes these differences very small\footnote{Strictly
  speaking, these contributions are needed to ensure that the final
  result is a real number. Indeed, when one picks poles like
  $p_0=\omega_{\imb p}+i\Gamma$ from the propagators and plugs them in
  the distribution functions, the latter become complex numbers. Their
  (small) imaginary part is canceled by the (small) contribution
  coming from the poles of $n_{_{F}}(r_0)-n_{_{F}}(p_0)$. For this
  approximation to be consistent, one must also neglect $i\Gamma$ in
  the hard argument of statistical weights.}. In other words, the only
important terms are those for which denominators like $P^2-M^2_\infty$
are small, and comparable to $p_0\Gamma$. This cannot happen if $p_0$
is an imaginary number of order $T$.

We are therefore left with the poles of the propagators themselves. At
this point, we obtain the following result:
\begin{eqnarray}
&&\hbox{\rm Im}\,\Pi^\mu{}_\mu(Q)\approx{1\over 2}
e^2 \int {{d^3 {\imb p}}\over{(2\pi)^3}}\sum\limits_{\eta=\pm 1}
\left[n_{_{F}}(q_0+\eta\omega_{\imb p})-n_{_{F}}(\eta\omega_{\imb p})\right]\;
\hbox{\rm Tr}({p_0=\eta\omega_{\imb p}})\nonumber\\
&&\qquad\qquad\times{\eta\over{2\omega_{\imb p}}}
\left[{1\over{(q_0+\eta\omega_{\imb p}+2i\Gamma)^2-\omega_{\imb r}^2}}-
{1\over{(q_0+\eta\omega_{\imb p}-2i\Gamma)^2-\omega_{\imb r}^2}}\right]\; .
\label{eq:1-loop-interm}
\end{eqnarray}
We can also note for later use that the result of this approximation
could have been obtained by starting from an expression slightly
different from Eq.~(\ref{eq:1-loop}):
\begin{eqnarray}
&&\hbox{\rm Im}\,\Pi^\mu{}_\mu(Q)\approx{1\over 2}
e^2 \int {{d^4 P}\over{(2\pi)^4}} 
\left[n_{_{F}}(r_0)-n_{_{F}}(p_0)\right]\;
\hbox{\rm Tr}\nonumber\\
&&\qquad\qquad\qquad \qquad\times 2\pi\epsilon(p_0)\delta(P^2-M^2_\infty)
\,
[{\imb\Delta}_{_{R}}^{2\Gamma}(R)
-{\imb\Delta}_{_{A}}^{2\Gamma}(R)]\ ,
\label{eq:1-loop-modifie}
\end{eqnarray}
where we denote
\begin{equation}
{\imb\Delta}_{_{R,A}}^{2\Gamma}(R)\equiv 
{1\over{(r_0\pm 2i\Gamma)^2-{\imb r}^2-M^2_\infty}}\; .
\end{equation}
In other words, if $\Gamma\ll T$, then we can as well put twice the
width on one of the two quark lines, and nothing in the other quark
line\footnote{This is where it is important to have a constant width.
  Indeed, some intermediate step relies on the cancellation
  $\Gamma(P)-\Gamma(R)=0$.}.

After Eq.~(\ref{eq:1-loop-interm}), the angular integration over
$d\Omega_{\imb p}$ is trivial. In the case where $\eta=+1$ (process
$q\to q\gamma$), we find:
\begin{eqnarray}
&&
{\rm Im}\,\Pi^\mu{}_\mu(Q)\approx {{e^2\Gamma}\over{2\pi^2 q_0}}
\int\limits_{0}^{+\infty}dp (q_0+p) [n_{_{F}}(q_0+p)-n_{_{F}}(p)]
\nonumber\\
&&\qquad\qquad\qquad\qquad\qquad\times
\ln\left(
{{(\omega_{\imb p}+p)^2q^2+\Gamma^2(q_0+p)^2}\over
{(\omega_{\imb p}-p)^2q^2+\Gamma^2(q_0+p)^2}}
\right)\; .
\end{eqnarray}
In particular, assuming for the sake of simplicity that $q\approx
q_0$, it is trivial to obtain the following asymptotic behaviors for
soft and hard photons:
\begin{eqnarray}
&&{\rm If\ \ }q_0 \ll T\; ,\qquad
{\rm Im}\,\Pi^\mu{}_\mu(Q)\sim e^2 \Gamma T\ln(1+4q_0^2/\Gamma^2)\; ,
\label{eq:1-loop-1a}\\
&&{\rm If\ \ }q_0 \gg T\; ,\qquad
{\rm Im}\,\Pi^\mu{}_\mu(Q)\sim e^2 \Gamma T \ln(T^2/\Gamma^2)\; .
\label{eq:1-loop-1b}
\end{eqnarray}
The main point is that these contributions are proportional to the
width and vanish when $\Gamma=0$, in agreement with a direct
calculation. As a side remark, one may note that for $\Gamma\sim g^2
T\ln(1/g)$ and soft photons ($q_0\sim gT$), Eq.~(\ref{eq:1-loop-1a})
is larger by a factor $1/g$ than the 1-loop HTL result, while in the
regime of Eq.~(\ref{eq:1-loop-1b}) it is of the same order. It is the
collision partners of the quarks that open up the phase-space (hard
quarks colliding in the medium can emit a photon, a process forbidden
for collisionless quarks) and make these contributions so large.

Another property of this result is that there is a suppression if
$q_0\ll T$, such that ${\rm Im}\,\Pi{}^\mu{}_\mu(Q)$ tends to 0 when
$q_0\to 0$. Equations similar to Eqs.~(\ref{eq:1-loop-1a}) and
(\ref{eq:1-loop-1b}) can be obtained when considering the contribution
$\eta=-1$ to Eq.~(\ref{eq:1-loop-modifie}).

\subsection{Cancellation with vertex corrections}
This suppression was interpreted in \cite{QuackH2} as the
manifestation of the LPM effect.  However, despite the suppression at
small $q_0$, the connection with the LPM effect is not clear in this
context. Indeed, the LPM effect is expected when the photon formation
time is larger than the quark mean free path, a condition which never
appears in \cite{QuackH2}, nor in the above 1-loop calculation. In
fact, following \cite{BlaizI1,BlaizI2}, we know that the propagator of
a quark close to its mass shell on which a width is resummed can be
evaluated by an eikonal resummation of the multiple collisions.  By
the same method, one can include in this resummation all the vertex
corrections. Indeed, one can check that the photon polarization tensor
at this level of approximation is proportional to (in space-time
coordinates)
\begin{equation}
\Pi{}^\mu{}^\nu(x,y)=e^2\int[dA^\mu] e^{iS[A]}{\rm Tr}\,[\gamma^\mu S(x,y|A)
\gamma^\nu S(y,x|A)]\; ,
\end{equation}
where $S[A]$ is the action of the gauge fields\footnote{This action
  does not play any role in the argument, and therefore can even
  include the HTL effective action for gluons.}, and where $S(x,y|A)$
is the propagator of a quark in the background field $A$. This formula
includes only one quark loop (in addition to the quark loops that may
have been resummed in the gluon propagators and vertices in $S[A]$),
and all orders in the gluon fields.  In the eikonal approximation, the
quark propagator has a very simple dependence on the field $A_\mu$
(which is a matrix $T^a A^a_\mu$ in QCD):
\begin{equation}
S(x,y|A)=S_0(x,y) {\cal P}\exp -ig\int\limits_{x^0}^{y^0} v_\mu A^\mu dt\; ,
\end{equation}
where $v_\mu$ is the 4-velocity of the quark, and $S_0(x,y)$ is the
free propagator of the quark. Now, if one is looking at photons of
very small invariant mass, the two quarks are nearly parallel in the
collinear limit (even if the photon is hard), so that the two quarks
have mostly the same $v$.  As a consequence, the two path ordered
exponentials cancel each other, and the product of the propagators
under the functional integral is
\begin{equation}
\gamma^\mu S(x,y|A)\gamma^\nu S(y,x|A)\approx \gamma^\mu S_0(x,y)\gamma^\nu S_0(y,x)\; .
\end{equation}
Therefore, at the level of approximation at which the resummed quark
propagator is calculated, the sum of {\sl all} the gluon loop
corrections is vanishing:
\setbox1=\hbox to 6cm{\resizebox*{6cm}{!}{\includegraphics{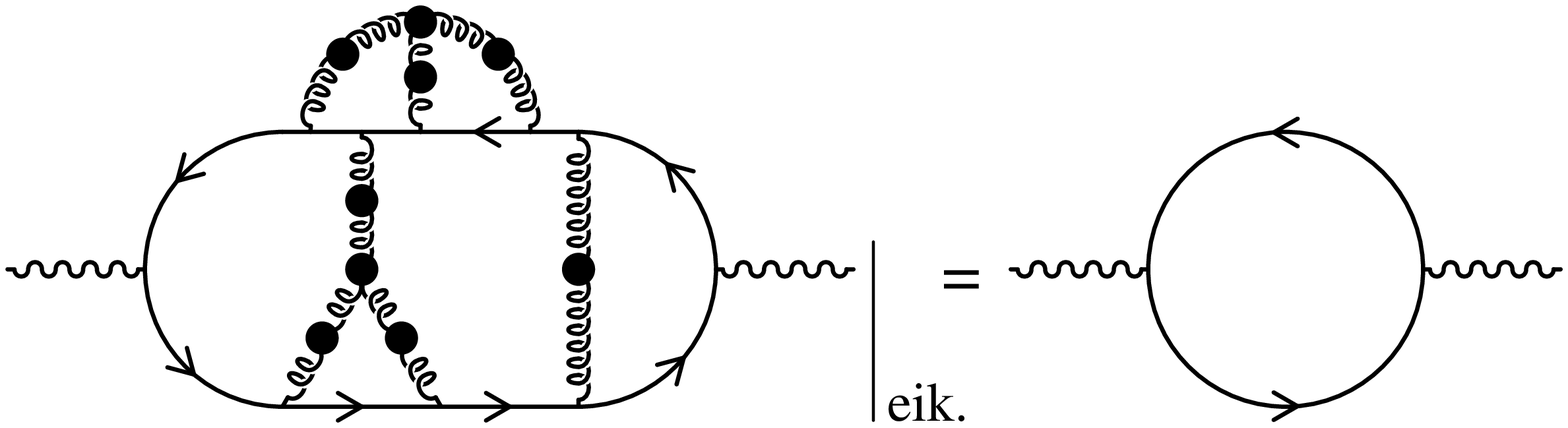}}}
\begin{equation}
  \sum\limits_{\rm all\ gluon\ corrections}\raise -5mm\box1\; .
\label{eq:eikon_approx}
\end{equation}
This result generalizes a result already known for QED established by
\cite{LebedS1,LebedS2,CarriKP1,CarriK1} for the subset of ladder
vertex corrections, and extended to all abelian topologies in
\cite{Petit1}, to the case where gluons are exchanged (i.e.  to non
abelian topologies). Indeed, we see that what makes gluons specific,
namely the fact that they can couple to each other, is hidden in the
action $S[A]$, which plays a passive role in the argument. 

This result is also closely related to the fact that there are no
$\gamma\gamma g\cdots g$ HTL vertices (see \cite{Bodek1} for an
interpretation of the cancellation found in
\cite{LebedS1,LebedS2,Petit1} as a consequence of the absence of HTL
vertices with $n>2$ photons), because the sum of all the eikonal
contributions can also be written as (when $Q$ is soft):
\setbox1=\hbox to
8cm{\resizebox*{8cm}{!}{\includegraphics{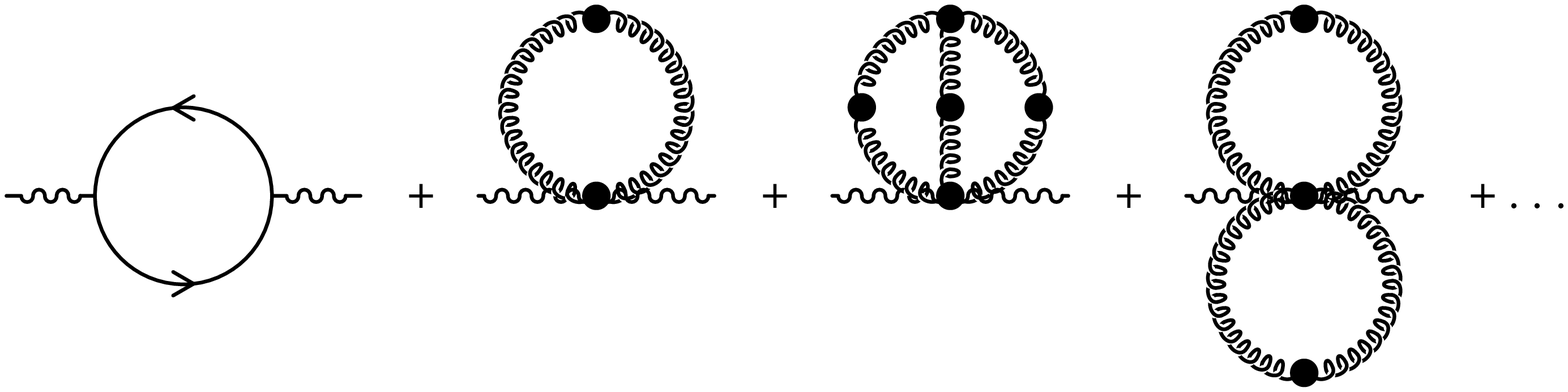}}\hfill}
\begin{equation}
  \Pi{}^\mu{}^\nu(Q)=\raise -9.5mm\box1\; ,
\label{eq:eikon_approx_HTL}
\end{equation}
where the sum is extended to all the possible ways to close the
external gluonic legs of the $\gamma\gamma g\cdots g$ vertices. If is
then obvious to see that the sum reduces to its first term if these
vertices are vanishing.  

One is therefore left with the bare 1-loop diagram, which does not
contributes to the imaginary part of the photon polarization tensor.
In other words, the results derived in Eqs.~(\ref{eq:1-loop-1a}) and
(\ref{eq:1-loop-1b}) as a warmup and that were claimed to be related
to the LPM effect in \cite{QuackH2}, are just artifacts with no
physical meaning.  In section \ref{sec:LPM} of the present paper, we
show where the LPM effect appears in thermal field theory.

Despite the fact that this contribution is not physical, one learns
two things from this calculation: (1) one has to keep terms beyond the
eikonal approximation in order to circumvent this cancellation, and
(2) the width may open the phase space to new processes.

\section{2-loop calculation}
\label{sec:2-loop-calcul}
The lesson from the previous section is that physical contributions
must be looked for beyond the eikonal approximation. This implies
inserting explicitly a gluon exchange with momentum $L$ in the
diagram, and not assuming that $L^2\ll 2P\cdot L$.

\begin{figure}[htbp]
\centerline{\resizebox*{!}{3.75cm}{\includegraphics{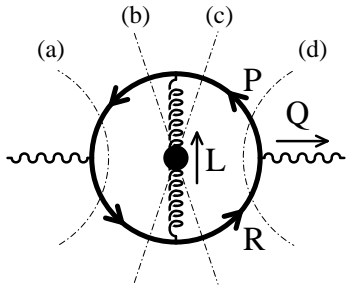}}}
\caption{\sl 2-loop vertex diagram. The dotted lines are the various cuts
  contributing to the imaginary part of the photon polarization
  tensor. The boldface quark line denotes the resummation of a width
  on the fermion propagator. The dot on the gluon indicates the
  resummation of HTLs on its propagator.}
\label{fig:diagrams}
\end{figure}

There are in principle two 2-loop topologies contributing to the
photon polarization tensor. However, it was found in \cite{AurenGKP2}
that only the topology correcting the $q\bar{q}\gamma$ vertex (see
Fig.~\ref{fig:diagrams}) is relevant in the region where the collinear
enhancement takes place. Since our purpose it to study how this
collinear enhancement is affected by the width $\Gamma$, we limit the
present study to the terms that were found important in
\cite{AurenGKP2}.

There is though one difference with the case $\Gamma=0$ that we must
take into account: with a zero width, processes corresponding to cuts
$(a)$ and $(d)$ are kinematically forbidden for small $Q^2$. As in
the previous section, turning on the width opens those production
channels, and we cannot disregard them {\sl a priori}.  This is not a
problem though from a purely technical point of view, because it turns
out that the sum of all the cuts has a simpler expression than each
individual piece\footnote{In the appendix \ref{app:cuts}, we show how
  one can obtain the difference between the cuts $(c)$ and $(d)$.
  Having already their sum, we can therefore reconstruct the two
  contributions separately.}.

If we select only terms that have the large Bose-Einstein factor
$n_{_{B}}(l_o)$, the contribution of Fig.~\ref{fig:diagrams} to the
imaginary part of the photon polarization tensor takes the following
simple form \cite{AurenGKP2}
\begin{eqnarray}
&&\hbox{\rm Im}\,\Pi^\mu{}_\mu(Q)\approx{1\over 2}
e^2 g^2 \int {{d^4 P}\over{(2\pi)^4}} 
\left[n_{_{F}}(r_0)-n_{_{F}}(p_0)\right]\nonumber\\
&&\qquad\qquad\qquad \qquad \times\int{{d^4 L}\over{(2\pi)^4}}  
n_{_{B}}(l_0) \rho_{_{T,L}}(L)
P^{\rho\sigma}_{_{T,L}}(L)\;
\hbox{\rm Tr}_{\rho\sigma}\nonumber\\
&&\qquad\qquad\qquad \qquad\times[{\imb\Delta}_{_{R}}(P){\imb\Delta}_{_{R}}(P+L)-{\imb\Delta}_{_{A}}(P){\imb\Delta}_{_{A}}(P+L)]
\nonumber\\
&&\qquad\qquad\qquad \qquad\times
[{\imb\Delta}_{_{R}}(R){\imb\Delta}_{_{R}}(R+L)-{\imb\Delta}_{_{A}}(R){\imb\Delta}_{_{A}}(R+L)]\ ,
\label{eq:cutvertex}
\end{eqnarray} 
where $\rho_{_{T,L}}(L)$ are the spectral functions of transverse and
longitudinal gluons, $P^{\rho\sigma}_{_{T,L}}(L)$ are the
corresponding projectors, and where ${\rm Tr}_{\rho\sigma}$ is the
result of the Dirac's trace for the quark loop.  The result of the
previous section requires that we keep in this Dirac's trace only
terms that do not appear in the eikonal approximation (i.e. which
includes soft corrections to the hard loop momentum). In the collinear
limit, the first non vanishing term (beyond eikonal approximation) is
\begin{equation}
{\rm Tr}_{\rho\sigma}\approx-8L^2(R_\rho R_\sigma+P_\rho P_\sigma)\; ,
\end{equation}
which turns out to be the same as the term found in \cite{AurenGKP2}.
Its contractions with the projectors are given by
\begin{eqnarray}
&&P^{\rho\sigma}_{_{T}}(L){\rm Tr}_{\rho\sigma}\approx -8L^2(r^2+p^2)(1-\cos^2\theta^\prime)\nonumber\\
&&P^{\rho\sigma}_{_{L}}(L){\rm Tr}_{\rho\sigma}\approx +8L^2(r^2+p^2)(1-\cos^2\theta^\prime)\; ,
\end{eqnarray}
where $\theta^\prime$ is the angle between the 3-vectors $\imb p$ and
$\imb l$. At this point, we have used the fact that we are looking at
collinearly enhanced terms, and consider that $\imb p$ and $\imb r$
are parallel (the only place where we do not do this approximation is
in the denominators which are very sensitive to the details of the
collinear sector).

In order now to perform the integral over $p_0$, we follow the method
of the previous section, and use the same approximations concerning
the statistical weights. In addition, we compute only the contribution
of cuts $(c)+(d)$, and multiply the result by an overall factor $2$ in
order to take into account the contribution of the cuts $(a)+(b)$.
Following the remark of the previous section, we can start directly
from the expression:
\begin{eqnarray}
&&\hbox{\rm Im}\,\Pi^\mu{}_\mu(Q)\approx{1\over 2}
e^2 g^2 \int {{d^4 P}\over{(2\pi)^4}} 
\left[n_{_{F}}(r_0)-n_{_{F}}(p_0)\right]\nonumber\\
&&\qquad\qquad \qquad \times\int{{d^4 L}\over{(2\pi)^4}}  
n_{_{B}}(l_0) \rho_{_{T,L}}(L)
P^{\rho\sigma}_{_{T,L}}(L)\;
\hbox{\rm Tr}_{\rho\sigma}\nonumber\\
&&\qquad\qquad \qquad\times 2\pi\epsilon(p_0)\delta(P^2-M^2_\infty)\,
{1\over{(P+L)^2-M^2_\infty}}\nonumber\\
&&\qquad\qquad\qquad\times
{\rm Disc}\left[{\imb \Delta}_{_{R}}^{2\Gamma}(R)\,
{\imb \Delta}_{_{R}}^{2\Gamma}(R+L)\right]\; ,
\label{eq:2-loop-start-simple}
\end{eqnarray} 
where we use the notation ${\rm Disc }f(\Gamma)\equiv
f(\Gamma)-f(-\Gamma)$. We first do the $p_0$ integration for free
thanks to the $\delta(P^2-M^2_\infty)$.  In this section, we consider
only the case of $p_0=+\omega_{\imb p}$ (bremsstrahlung of a quark) in
order to keep the calculation compact. 
\begin{figure}[htbp]
\centerline{\resizebox*{!}{2.5cm}{\includegraphics{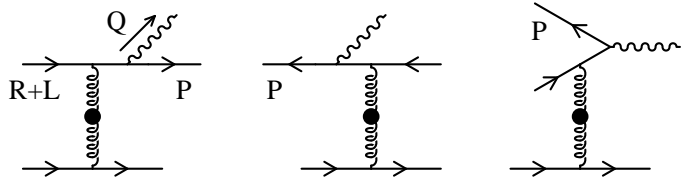}}}
\caption{\sl Different processes contained in the 2-loop
  diagram of Fig.~\ref{fig:diagrams}. Left: $p_0=+\omega_{\imb p}>0$.
  The corresponding process is the bremsstrahlung of a quark. Middle:
  $p_0=-\omega_{\imb p}<0$ and $\omega_{\imb p}> q_0$ so that
  $r_0=p_0+q_0$ (and also $r_0+l_0$) is negative as well. This region
  contains the bremsstrahlung of an antiquark. Right:
  $p_0=-\omega_{\imb p}<0$ and $\omega_{\imb p}< q_0$ so that
  $r_0=p_0+q_0$ (and also $r_0+l_0$) is positive. The corresponding
  process is the annihilation of an antiquark with an off-shell
  quark.}
\label{fig:processes}
\end{figure}
The contribution of $p_0<0$ is identical if $q_0\ll T$, but is
different if $q_0$ is large. In the latter case, the corresponding
process is a $q^*\bar{q}$ annihilation ($q^*$ denotes a quark placed
off-shell by a scattering) instead of bremsstrahlung (see
Fig.~\ref{fig:processes}), and is considered in more detail in section
\ref{sec:annihilation}.

Then, it happens that the angular integral over the direction
$\Omega_{\imb l}$ of the 3-vector $\imb l$ can be done analytically in
a rather simple way. We have to perform an integral
like\footnote{Strictly speaking, we have also a factor of
  $1-\cos^2\theta^\prime$ in the numerator that depends on
  $\Omega_{\imb l}$. This factor can be taken into account
  analytically in the angular integral, the price to pay being more
  cumbersome expressions. One can however make the following
  simplification: if the width is small and if we use the collinear
  approximation for the numerators, then this angle is approximately
  given by $\cos\theta^\prime\approx l_0/l$, and does not play any
  role in the angular integrals.}
\begin{equation}
I_{_{L}}\equiv\int{{d\Omega_{\imb l}}\over{4\pi}} {1\over{2L\cdot P+L^2}}\; 
{1\over{2L\cdot R +L^2 + 2P\cdot Q+Q^2+4i r_0\Gamma}}\; ,
\end{equation}
which can be rewritten as
\begin{equation}
I_{_{L}}=\int{{d\Omega_{\imb l}}\over{4\pi}} {1\over{2\widehat{L}\cdot A}}\; 
{1\over{2\widehat{L}\cdot B}}\; ,
\end{equation}
where $\widehat{L}\equiv(1,\hat{\imb l})$ provided we introduce the
fictitious ``4-vectors''
\begin{eqnarray}
&&A\equiv(p_0l_0+{L^2\over 2},l{\imb p})\nonumber\\
&&B\equiv(r_0l_0+{L^2\over 2}+P\cdot Q+{Q^2\over 2}+2ir_0\Gamma,l{\imb r})\; .
\end{eqnarray}
The advantage of rewriting $I_{_{L}}$ like this lies in the fact that
the last integral is known in closed form\footnote{$I_{_{L}}$ is an
  analytic function of its arguments. The two possible choices for the
  square root of the complex number $\Delta$ lead to the same
  result.}:
\begin{equation}
I_{_{L}}= {1\over{8\sqrt{\Delta}}}\left[\ln({A\cdot B+\sqrt{\Delta}})
-\ln({A\cdot B-\sqrt{\Delta}})\right]\; ,
\label{eq:IL-fin}
\end{equation}
where $\Delta\equiv (A\cdot B)^2-A^2 B^2$.  We need now to evaluate
the three quantities $A^2$, $B^2$ and $A\cdot B$, for which we will
also use approximations based on the fact that $L$ is soft while the
other momenta are hard. We obtain first a rather simple, and by now
very familiar \cite{AurenGKP2}, expression
for $\Delta$:
\begin{equation}
\Delta\approx p^4 q_0^2 l^2\left[
\Big(1-\cos\theta+{{M^2_{\rm eff}}\over{2p^2}}+{{L^2}\over{2p^2}}\Big)^2
-{{L^2}\over{p^2}}{{M^2_{\rm eff}}\over{p^2}}
\right]\; ,
\end{equation}
where $\theta$ is the angle between the 3-vectors $\imb p$ and $\imb
q$, and with
\begin{equation}
M^2_{\rm eff}\equiv M^2_\infty + {{Q^2}\over{q_0^2}}p(p+q_0) +
4i {\Gamma\over {q_0}} p(p+q_0)\; .
\label{eq:Meff}
\end{equation}
Therefore, apart from the fact that the effective mass $M^2_{\rm eff}$
now gets an imaginary part coming from the width (and has been
extended to hold for hard $q_0$ as well), this quantity is nothing
but the denominator appearing in Eq.~(39) of \cite{AurenGKP2}.  

Since $A\cdot B\approx pr L^2 \gg\sqrt{\Delta}$ and since the
bremsstrahlung lives in the region where $L^2<0$, the difference of
the two logarithms in Eq.~(\ref{eq:IL-fin}) is just the discontinuity
of the logarithm across its branch cut, which gives
\begin{equation}
\ln({A\cdot B+\sqrt{\Delta}})
-\ln({A\cdot B-\sqrt{\Delta}})
\approx 2i\pi\epsilon(\Gamma)\; .
\end{equation}
Therefore, in presence of a width $\Gamma$, the contribution of
bremsstrahlung to the photon polarization tensor is
\begin{eqnarray}
&&{\rm Im}{}^\mu{}_\mu(Q)\approx {{e^2 g^2}\over{32\pi^4}}
{T\over{q_0^2}}\int\limits_{0}^{+\infty}dp\; {{p^2+(p+q_0)^2}
\over{p^2}} [n_{_{F}}(p+q_0)-n_{_{F}}(p)]
\nonumber\\
&&\qquad\times
\int{{dl_0}\over{l_0}}\int l^3 dl\; [\rho_{_{T}}(l_0,l)-\rho_{_{L}}(l_0,l)] 
\Big[1-\Big({{l_0}\over{l}}\Big)^2\Big]
\nonumber\\
&&\qquad\times
{\rm Disc}\int\limits_{0}^{2}{{\epsilon(\Gamma)\;du}\over
{\Big[
u+{{M^2_{\rm eff}}\over{2p^2}}
\Big]
\Big[
\Big(
u+{{M^2_{\rm eff}}\over{2p^2}}+{{L^2}\over{2p^2}}
\Big)^2-{{L^2}\over{p^2}}{{M^2_{\rm eff}}\over{p^2}}
\Big]^{1/2}
}}\; ,
\label{eq:2-loop-interm}
\end{eqnarray}
where $u\equiv1-\cos\theta$. At this stage, this expression is
formally similar to the one found for $\Gamma=0$ and reproduces known
results in this limit. Indeed, when $\Gamma=0$, then $M^2_{\rm eff}$
is a real number, and taking the discontinuity just gives a factor of
$2$ (do not forget the $\epsilon(\Gamma)$ in the numerator).  It is
then possible to use the very same sequence of changes of variables as
in \cite{AurenGKP2} to transform the integral over $u$, and write
${\rm Im}\Pi{}^\mu{}_\mu(Q)$ as\footnote{We used the fact that the
  spectral functions $\rho_{_{T,L}}$ have a simple expression in the
  space-like region:
\begin{equation}
\rho_{_{T,L}}(L)={{-2\,{\rm Im}\,\Pi_{_{T,L}}^{^{HTL}}(L)}
\over{(L^2-{\rm Re}\,\Pi_{_{T,L}}^{^{HTL}}(L))^2+
({\rm Im}\,\Pi_{_{T,L}}^{^{HTL}})^2}}\; .
\end{equation}}
\begin{eqnarray}
&&
{\rm Im}\Pi{}^\mu{}_\mu(Q)\approx {{2e^2 g^2}\over{\pi^4}}{T\over{q_0^2}}
\int\limits_{0}^{+\infty} dp\; {{p^2+(p+q_0)^2}\over 2}
[n_{_{F}}(p+q_0)-n_{_{F}}(p)]
\nonumber\\
&&
\qquad\qquad\qquad\qquad\times
\sum\limits_{m=T,L}
\int\limits_{0}^{1}{{dx}\over x}\;\int\limits_{0}^{+\infty}dw\;
{{\left|\widetilde{I}_{m}\right|\,K(w,\widehat{\Gamma})}
\over
{(w+\widetilde{R}_{m})^2+(\widetilde{I}_{m})^2}}
\; ,
\label{eq:2-loop-final}
\end{eqnarray}
where we denote:
\begin{eqnarray}
&&w\equiv {{-L^2}\over{{\rm Re}\,M^2_{\rm eff}}}\; ,\qquad
x\equiv{{l_0}\over{l}}\; , \qquad
\widehat{\Gamma}\equiv {{{\rm Im}\,M^2_{\rm eff}}\over{{\rm
  Re}\,M^2_{\rm eff}}}\nonumber\\
&& \widetilde{I}_{_{T,L}}\equiv
{{{\rm Im}\,\Pi^{^{HTL}}_{_{T,L}}}\over{{\rm Re}\,M^2_{\rm eff}}}\; ,
\qquad
\widetilde{R}_{_{T,L}}\equiv
{{{\rm Re}\,\Pi^{^{HTL}}_{_{T,L}}}\over{{\rm Re}\,M^2_{\rm eff}}}\; ,
\label{eq:2-loop-variables}
\end{eqnarray}
and where the function $K$ comes from the integral over $u$ and is
defined as\footnote{One can go from the variable $u$ to $y$ by the
  following transformations
\begin{equation}
{2p^2u}\equiv M^2_{\rm eff}[w(1-z)+z^{-1}-1]\; ,
\qquad{\rm and\ \ }y\equiv 4(z-z^2)\; .
\end{equation}}
\begin{eqnarray}
&&
K(w,\widehat{\Gamma})\equiv {1\over 2}\int\limits_{0}^{1}{{dy}\over{\sqrt{1-y}}}
{{y+4/w}\over{(y+4/w)^2+(4\widehat{\Gamma}/w)^2}}
\nonumber\\
&&\qquad={1\over{4(\alpha^2+\beta^2)}}\Big\{
\alpha\ln\Big(
{{(1+\alpha)^2+\beta^2}\over{(1-\alpha)^2+\beta^2}}
\Big)
\nonumber\\
&&\qquad\quad
-2\beta\Big[
\arctan\Big(
{{\alpha+1}\over{\beta}}
\Big)
-\arctan\Big(
{{\alpha-1}\over{\beta}}
\Big)
\Big]\Big\}
\;
\end{eqnarray}
with $\alpha+i\beta$ a square root\footnote{Explicitly, we
  have:
\begin{equation}
\alpha,\beta=\sqrt{{1\over 2}\Big[
\sqrt{((4+w)/w)^2+(4\widehat{\Gamma}/w)^2}\pm((4+w)/w)
\Big]}\; .
\end{equation}
} of the complex number $(4+w)/w+4i\widehat{\Gamma}/w $. This function
$K(w,\widehat{\Gamma})$ is the generalization to the case of a non
vanishing width of the factor $\sqrt{w/(w+4)}\tanh^{-1}\sqrt{w/(w+4)}$
appearing in Eq.~(89) of \cite{AurenGKP2}.  In the limit of vanishing
width $(\widehat{\Gamma}\to 0)$, we recover the results of
\cite{AurenGKP2}.  Eq.~(\ref{eq:2-loop-final}) cannot be further
simplified analytically (except in some limiting cases), and must be
evaluated numerically.

\section{Discussion}
\label{sec:2-loop-discussion}
\subsection{Modification of the emission spectrum}
By inspecting our final expression given in
Eq.~(\ref{eq:2-loop-final}), the first thing we notice is the very
strong similarity with the same result in the absence of the width.
Only the function $K(w,\widehat{\Gamma})$ contains the width, in the
form of the dimensionless ratio $\widehat{\Gamma}$. When this ratio is
small, the width has a negligeable effect while on the contrary if
$\widehat{\Gamma}\gg 1$ then the width plays a dominant role.

\begin{figure}[htbp]
\centerline{\rotatebox{-90}{\resizebox*{!}{7cm}{\includegraphics{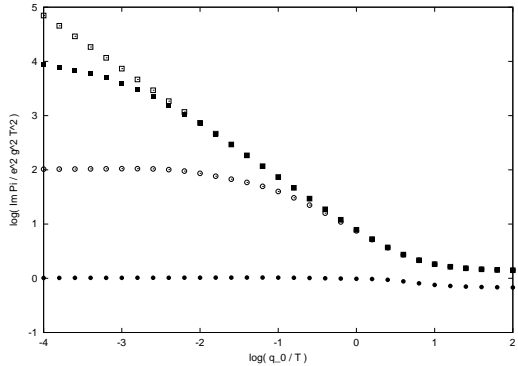}}}}
\caption{\sl Effect of the width on the bremsstrahlung as a function 
  of $q_0/T$ (for $Q^2=0$). The various curves correspond to different
  values of the width $\Gamma$. From top to bottom, the ratio $\Gamma
  T/M^2_\infty$ (which corresponds to $\widehat{\Gamma}/4$ when
  $q_0\gg T$) takes the values $10^{-6}$, $10^{-4}$, $10^{-2}$ and
  $1$.}
\label{fig:width-effect}
\end{figure}

A first simple conclusion is obtained by noticing that the width
arrives in $M^2_{\rm eff}$ via the combination $\Gamma p(p+q_0)/q_0$
which becomes large at small $q_0$. Therefore, the effect of the width
is more important at small $q_0$. This property is illustrated by the
plot of Fig.~\ref{fig:width-effect}, which shows ${\rm
  Im}\,\Pi{}^\mu{}_\mu$ (obtained numerically from
Eq.~(\ref{eq:2-loop-final})) as a function of the photon energy $q_0$
(the invariant mass $Q^2$ is kept zero in this plot), for different
values of the width $\Gamma$.
On this plot, one can see that the $q_0^{-1}$ behavior of
bremsstrahlung at small $q_0$ is modified in presence of a width.
Instead of that, one reaches a plateau\footnote{A similar behavior has
  already been observed in a very different calculation by Weldon in
  \cite{Weldo6} (see Fig.~1 of \cite{Weldo6}). However, the
  resummations considered in this paper deal with the possibility to
  emit more than one photon, and affect the spectrum only below the
  scale $q_0\sim \alpha_{_{EM}}T$, much smaller than the scale at
  which $\Gamma\sim g^2 T\ln(1/g)$ starts playing a role. The effect
  we are considering here appears at much larger photon energies (even
  including hard photons if $\Gamma$ is large enough), since it is
  related to resummations of gluons and thus involves the strong
  coupling constant $\alpha_{_{S}}\gg\alpha_{_{EM}}$.} when $q_0\to
0$. By comparing the second and third curves, we see that the value of
$q_0$ at which we reach this plateau varies by a factor $10^2$ when
$\Gamma$ varies by the same factor. This is a consequence of the fact
that the width enters in the result via the ratio $\Gamma/q_0$ for
small $q_0$.  Moreover, the value of the plateau is proportional to
$1/\Gamma$ (this can be deduced from the plot since $\Gamma$ increases
by a factor $10^2$ between two consecutive curves).  This leads to the
conjecture that the rate has a very simple $\Gamma$ dependence in the
region where the width is dominant.  Additionally, when $\Gamma$
increases, the extension of this plateau increases as well. In
particular, for a large enough width, even the region of hard photons
is modified.

\subsection{Region where the width is important}
\begin{figure}[htbp]
\centerline{\rotatebox{-90}{\resizebox*{!}{7cm}{\includegraphics{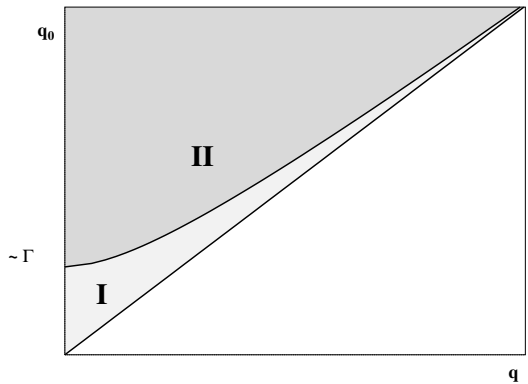}}}}
\caption{\sl Boundary obtained from the condition 
  $\widehat{\Gamma}=1$. In region I, the width is the dominant
  regulator of collinear singularities. In addition, the shape of the
  emitted spectrum is modified by the width. In region II, the width
  is only a sub-dominant correction.}
\label{fig:boundary}
\end{figure}

If one wants to go beyond this qualitative statement, and get a better
idea of the region (in the $(q,q_0)$ plane) where the width is
important, one must study the ratio $\widehat{\Gamma}$. Indeed, the
curve defined by the equation $\widehat{\Gamma}=1$ is precisely the
curve on which the effect of the width is comparable to the effect of
$M^2_\infty$ and $Q^2$ (see Fig.~\ref{fig:boundary}). Above this curve
(region II of Fig.~\ref{fig:boundary}), the width is an irrelevant
parameter, and below this curve (region I -- low invariant mass
photons) we have the region where the width is the dominant collinear
regulator.

At first sight, the region where the width is important looks very
similar to the region where higher loop corrections are IR-sensitive
to the scale $g^2T$ and must be resummed, as found in \cite{AurenGZ1}.
We recall that in that work we considered the infrared structure of
higher loop diagrams contributing to thermal photon production. It was
shown that compensations between different cuts cancel all unscreened
infrared divergences, and that the remaining terms are sensitive to
gluon momenta down to a $Q$-dependent cutoff. In addition, we compared
this cutoff with the scale $\mu\sim g^2T$ of the magnetic mass, and
concluded that higher loop diagrams would be sensitive to the magnetic
mass for a small enough photon invariant mass (see the figure 5 of
\cite{AurenGZ1}).

This similarity comes from the fact that line in the $(q,q_0)$ plane
where $\Gamma$ starts to be important is given by the equation:
\begin{equation}
2\Gamma\sim {{q_0}\over{2}}{{{\rm Re}\,M^2_{\rm eff}}\over{p(p+q_0)}}\; .
\label{eq:condition-on-gamma}
\end{equation}
while the line on which the magnetic mass $\mu$ becomes important has
the following equation:
\begin{equation}
\mu\sim {{q_0}\over{2}}{{{\rm Re}\,M^2_{\rm eff}}\over{p(p+q_0)}}\; .
\label{eq:condition-on-mu}
\end{equation}
In other words, the two lines are defined by comparing a common
momentum scale alternatively with the width of the quarks and with the
magnetic mass. The physical interpretation of the momentum scale
appearing in the right hand side of Eqs.~(\ref{eq:condition-on-gamma})
and (\ref{eq:condition-on-mu}) will be given in the next section.  The
two boundaries are therefore very similar in QCD because $\Gamma\sim
g^2T\ln(1/g)$ and $\mu\sim g^2T$ are not very different.  But on the
other hand, $\Gamma$ and $\mu$ have very different
meanings\footnote{In particular, the fact that $\Gamma$ and $\mu$ are
  very close is specific to QCD. In QED, one would have $\mu=0$ and
  $\Gamma\sim e^2 T\ln(1/e)$.}, and we may expect rather different
interpretations for the two conditions
Eqs.~(\ref{eq:condition-on-gamma}) and (\ref{eq:condition-on-mu}). In
the next section, we show that Eq.~(\ref{eq:condition-on-gamma}) is
closely related to the LPM effect.

There is also a more technical argument suggesting the difference of
Eqs.~(\ref{eq:condition-on-gamma}) and (\ref{eq:condition-on-mu}),
which becomes clear by examining how the width $\Gamma$ would appear
in perturbation theory. For that purpose, let us insert a self-energy
correction on the leg of momentum $R$ in the two-loop diagram of
Fig.~\ref{fig:diagrams}. Since we are dealing with the width
perturbatively here, the quark propagators contain only the asymptotic
mass $M_\infty$. By a crude power counting, we can estimate that this
insertion brings the following extra factor\footnote{A self-energy
  correction of the type considered here modifies the real-part of the
  pole of the propagator by a mass-shift $\delta M^2_\infty$,
  negligible compared to $M^2_\infty$ arising from HTL corrections.}
\begin{equation}
{{ir^0\Gamma}\over{R^2-M^2_\infty}}\sim {1\over{{\rm Re}\,M^2_{\rm eff}}}
{{i\Gamma p (p+q_0)}\over{q_0}}\; .
\end{equation}
We can already see that a self-energy insertion increases the strength
of the potential collinear divergences by bringing an additional
denominator $R^2-M^2_\infty$. On the other hand, this insertion does
not modify the infrared properties of this diagram. By summing over
the number of such insertions from $0$ to $+\infty$, we get a factor
\begin{equation}
{{{\rm Re}\,M^2_{\rm eff}}\over{{\rm Re}\,M^2_{\rm eff}\oplus i\Gamma p(p+q_0)/q_0}}\; .
\end{equation}
Therefore, the effect of such a resummation is to substitute ${\rm
  Re}\,M^2_{\rm eff}$ by ${\rm Re}\,M^2_{\rm eff}\oplus i\Gamma
p(p+q_0)/q_0$ in the factor $T^2/ M^2_{\rm eff}$ of collinear
enhancement. This is precisely what has been observed in the more
rigorous calculation of section \ref{sec:2-loop-calcul}.  In addition,
this simple argument demonstrates clearly that the mode of action of
width insertions is to affect the collinear sector, leaving unmodified
the infrared sector.  On the contrary, the possibility to have a
sensitivity to $\mu$ found in \cite{AurenGZ1} is related to infrared
singularities due exclusively to transverse gluons\footnote{In
  particular, longitudinal as well as transverse gluons contribute to
  $\Gamma$. There is no contradiction with \cite{AurenGZ1} which found
  only the transverse gluons to be important, since the statement of
  \cite{AurenGZ1} was about infrared singularities.}.

\subsection{Limit of dominant width}
In region I, where the width becomes the dominant regulator, the
ratio $\widehat{\Gamma}$ is large, which enables us to make some
additional approximations. In particular, we can perform very simply
the integrals over $w$ and $y$ in Eq.~(\ref{eq:2-loop-final}). Indeed,
we can first note that $\widehat{\Gamma}$ sets the order of magnitude of the
variable $w$. As a consequence, typical values of $w$ are large, and
we can neglect the corrections $\widetilde{R}_{_{T,L}}$ and
$\widetilde{I}_{_{T,L}}$ in the denominator of
Eq.~(\ref{eq:2-loop-final}), as well as $1/w$ in front of $y$.
Therefore, this equation becomes
\begin{eqnarray}
&&
{\rm Im}\Pi{}^\mu{}_\mu(Q)\approx {{2e^2 g^2}\over{\pi^4}}{T\over{q_0^2}}
\int\limits_{0}^{+\infty} dp\; {{p^2+(p+q_0)^2}\over 2}
[n_{_{F}}(p+q_0)-n_{_{F}}(p)]
\nonumber\\
&&
\qquad\qquad\times
\left[\sum\limits_{m=T,L}
\int\limits_{0}^{1}{{dx}\over x}
\left|\widetilde{I}_{m}\right|\right]
\;\left[\int\limits_{0}^{+\infty}{{dw}\over{w^2}}\;
{1\over 2}\int\limits_{0}^{1}{{dy}\over{\sqrt{1-y}}}
{y\over{y^2+(4\widehat{\Gamma}/w)^2}}\right]
\nonumber\\
&&={{2e^2 g^2}\over{\pi^4}}{T\over{q_0^2}}
\!\!\int\limits_{0}^{+\infty} \!\!\!dp\, {{p^2+(p+q_0)^2}\over 2}
[n_{_{F}}(p+q_0)-n_{_{F}}(p)]\times
{{3\pi m^2_{\rm g}}\over{2 {\rm Re}\,M^2_{\rm eff}}}
{{\pi}\over{4\widehat{\Gamma}}}
\label{eq:2-loop-final-large-width}
\end{eqnarray}
where $m_{\rm g}\sim gT$ is the gluon thermal mass coming from the
prefactor of $\widetilde{I}_{_{T,L}}$.  We see that this expression is
of the form
\begin{equation}
{\rm Im}\Pi{}^\mu{}_\mu(Q)\approx e^2 g^4 {{T^3}\over{\Gamma}} 
\Phi\Big({{q_0}\over T}\Big)\; ,
\label{eq:2-loop-large-gamma}
\end{equation}
where $\Phi$ is a function independent of $\Gamma$ giving the shape of
the photon spectrum. Therefore, in the region where the width
dominates, the spectrum scales as $\Gamma^{-1}$. One notes that the
combination $\widehat{\Gamma}\,{\rm Re}\,M^2_{\rm eff}$ is
proportional to $\Gamma p(p+q_0)/q_0$, and therefore the shape of the
spectrum is independent of $Q^2$ and $M^2_\infty$. Furthermore, for
$q_0\ll T$, it is independent of $q_0$ (see
Fig.~\ref{fig:width-effect}).  

It may be interesting to compare in two extreme cases ($q_0$ soft and
$q_0\gg T$) the result of Eq.~(\ref{eq:2-loop-large-gamma}) with those
obtained with a vanishing width. In the soft $q_0$ case, the result
when $\Gamma=0$ was obtained in \cite{AurenGKP2}:
\begin{equation}
{\rm Im}\,\Pi{}^\mu{}_\mu(Q)\sim e^2 g^4 {{T^3}\over{q_0}} {1\over{g^2}}\; ,
\end{equation}
where we have explicitly isolated the factor $T^2/M^2_\infty \sim
1/g^2$ that comes from the collinear enhancement, while we have now:
\begin{equation}
{\rm Im}\,\Pi{}^\mu{}_\mu(Q)\sim e^2 g^4 {{T^3}\over{q_0}} 
{{q_0}\over\Gamma}\; .
\end{equation}
The factor $q_0/\Gamma$ is in fact the new enhancement factor coming
from $T^2/{\rm Im}\,M^2_{\rm eff}$. Therefore, the two results differ
only by the nature of the factor of collinear enhancement. For hard
photon bremsstrahlung, we had on the other hand \cite{AurenGKZ1}
\begin{equation}
{\rm Im}\,\Pi{}^\mu{}_\mu(Q)\sim e^2 g^4 T^2 {1\over{g^2}}\; ,
\label{eq:hard-no-width}
\end{equation}
which becomes now
\begin{equation}
{\rm Im}\,\Pi{}^\mu{}_\mu(Q)\sim e^2 g^4 T^2 {{T}\over{\Gamma}}\; .
\label{eq:hard-width}
\end{equation}
Again, the two expressions differ only by the collinear enhancement
factor: $T^2/M^2_\infty \sim 1/g^2$ if $\Gamma=0$ instead of $T^2/{\rm
  Im}\,M^2_{\rm eff}\sim T/\Gamma$ when the width is dominant.  The
numerical prefactors not written in Eqs.~(\ref{eq:hard-no-width}) and
(\ref{eq:hard-width}) could however be quite different since they
reflect a different physics.

The $\Gamma^{-1}$ scaling law in Eq.~(\ref{eq:2-loop-large-gamma}) has
also a very simple physical interpretation related to the fact that
$\Gamma^{-1}$ is the mean free path of the quark in the medium.  This
result just tells us that the photon rate is proportional to the mean
free path of the quark that emits the photon. In other words, quarks
colliding very frequently do not have enough time to emit photons.
This picture is at the basis of the Landau-Pomeranchuk-Migdal effect,
that we discuss in the next section.

\section{Connection with the LPM effect}
\label{sec:LPM}
\subsection{Generalities on the LPM effect}
The non-perturbative region found in the present paper can also be
interpreted in a much more physical way, which seems to indicate that
the region I in Fig.~\ref{fig:boundary} is the region where the
Landau-Pomeranchuk-Migdal
\cite{LandaP1,LandaP2,Migda1,BaierDMPS1,Zakha2} effect modifies photon
production by a plasma.

Let us first recall the condition for the LPM effect, by using a very
heuristic argument valid for real photon production. In the
bremsstrahlung process of photon production, the gluon exchanged during
the scattering has a minimal momentum given by \cite{Migda1}
\begin{equation}
l_{\rm min}\approx {{q_0}\over{2}} {{M^2_\infty}\over{p(p+q_0)}}\; ,
\end{equation}
where $p$ is the momentum of the quark.  The inverse of this momentum
transfer defines the ``length'' on which the photon is emitted, and
for this reason is called the ``coherence length'' (denoted
$\lambda_{\rm coh}$ in the following) for the production of this
photon. It is also interpreted as the ``formation time'' of the
photon.  If $\lambda_{\rm coh}$ is much smaller that the typical
distance between two consecutive scatterings (proportional to
$1/\Gamma$), then the photon rate is not affected by multiple
scatterings. In other words, successive scatterings can be considered
as independent. On the contrary, if the coherence length is larger
than the mean free path, then successive scatterings are not
independent anymore. This is the LPM effect. It is also possible to
recast the previous condition as a comparison between the ``formation
time'' and the mean free path of the quark in a very suggestive way: the
photon must be produced before the quark scatters off another parton.

\subsection{LPM effect in thermal field theory}
Let us now show that the previous discussion arises automatically in
the thermal field theory approach.  In fact, there is in the
calculation presented before a quantity very similar to the coherence
length discussed in the semi-classical treatment of the LPM effect,
namely,
\begin{equation}
\lambda_{\rm coh}^{-1}\sim
 {{q_0}\over{2}} {{{\rm Re}\,M^2_{\rm eff}}
\over{p(p+q_0)}}\; . 
\end{equation}
This value agrees with the one of \cite{LandaP1,LandaP2,Migda1} when
$Q^2=0$ since it comes from kinematics, and generalizes it to the case
where the invariant mass $Q^2$ is non vanishing. Additionally, this
quantity was found in \cite{AurenGZ1} to be the lower bound for the
momentum of exchanged gluons.

It is now possible to reformulate the condition $\widehat{\Gamma}=1$
in more physical terms. Indeed, this condition can be rewritten as
\begin{equation}
\lambda_{\rm mean}\sim \Gamma^{-1}\sim\lambda_{\rm coh}\; ,
\label{eq:LPM-condition}
\end{equation}
where $\lambda_{\rm mean}$ is the mean free path of the quark in the
plasma, and the region where the width dominates corresponds to the
condition $\lambda_{\rm mean} < \lambda_{\rm coh}$. At this point, the
connection with the LPM effect is obvious: the region we found to be
non-perturbative due to the width of the quarks is also the region
where the LPM effect matters.

This discussion on the condition for the LPM effect to occur is in
fact summarized in the expression for $M^2_{\rm eff}$ introduced in
Eq.~(\ref{eq:Meff}). This formula can be written very elegantly as:
\begin{equation}
M^2_{\rm eff}={{2p(p+q_0)}\over{q_0}}\Big[
{1\over{\lambda_{\rm coh}}}+ {i\over{\lambda_{\rm mean}}}
\Big]\; ,
\label{eq:M_eff-compact}
\end{equation}
and the physical condition for the LPM effect\footnote{In
  \cite{RaffeS1} was investigated the influence of the finite mean free path
  of nucleons on axion (or neutrino pair) production by a supernova.
  The connection between the width $\Gamma$ and the LPM effect is also
  mentioned in this paper. However, the condition given by
  Eq.~(\ref{eq:LPM-condition}) does not come out from the formalism of
  \cite{RaffeS1} (it seems that the approximations made in
  \cite{RaffeS1} for the bremsstrahlung do not enable one to track the
  coherence length, so that only $\Gamma$ appears in the final result).
  What we have shown in the present section is that a careful
  calculation of the bremsstrahlung in thermal field theory generates
  automatically both terms of the comparison, through the effective
  mass given in Eq.~(\ref{eq:M_eff-compact}).} to be relevant is
mathematically expressed by the dominance of ${\rm Im}\,M^2_{\rm eff}$
over ${\rm Re}\,M^2_{\rm eff}$. Conversely, when ${\rm Re}\,M^2_{\rm
  eff} > {\rm Im}\,M^2_{\rm eff}$, the perturbative approach is valid.
As a side remark, let us note that the quantity $M^2_{\rm eff}$, which
controls the physics of bremsstrahlung, combines in a very simple way
three soft scales of the problem: $Q^2$, $M^2_\infty$ and $\Gamma$.
It is also possible to give a very simple physical interpretation of
the ratio $\widehat{\Gamma}$ which contains all the $\Gamma$
dependence of the final result (Eq.~(\ref{eq:2-loop-final})).
Indeed, one can rewrite this ratio as $\widehat{\Gamma}=\lambda_{\rm
  coh}/\lambda_{\rm mean}$, which is nothing but the typical number of
coherent scatterings necessary to produce a photon.

The analogy with the standard treatment of the LPM effect is only
partial though. Indeed, our estimate of the effect of the finite
mean free path of the quarks is too crude (because of the constant width,
and because of the missing vertex corrections) to be reliable from a
quantitative perspective. The trend found here (suppression of the
photon spectrum at small $q_0$) is however expected to be a solid
prediction, and is in agreement with what is usually found from the
LPM effect. 

The main difference with the usual calculation of the LPM effect
comes from the treatment of scattering centers: one usually assumes a
fast moving charged particle going through a medium of ``cold''
(static) scattering centers. In the case we are considering here, both
the particle emitting the photon and the scattering centers are
thermalized particles, of comparable momenta. The approximations made
in \cite{LandaP1,LandaP2,Migda1,BaierDMPS1,Zakha2} do not apply to
this situation, and it is not clear whether a quantitative agreement
is to be expected at all. In particular, the static approximation
selects only Debye shielded longitudinal gluon exchanges, while the
width (the resummation of which introduces multiple scatterings in our
approach) of the thermalized quark receives contributions from
transverse gluons as well.  Contrary to
\cite{LandaP1,LandaP2,Migda1,BaierDMPS1,Zakha2}, transverse gluon
exchanges are very important in thermal field theory, and matter for
the LPM effect as well. 

There is a major difference between the LPM effect from multiple
longitudinal gluon exchanges and multiple transverse gluon exchanges,
which can be readily seen when one compares the respective ranges of
electric and magnetic fields to the mean free path of the quark.
Indeed, since $\lambda_{\rm mean} \gg m_{_{D}}^{-1}\sim (gT)^{-1}$,
the mean free path is much larger than the range of Debye screened
electric fields.  As a consequence, successive exchanges of
longitudinal gluons are independent: they correspond to scatterings
off different partons. On the contrary, we have $\mu^{-1} >
\lambda_{\rm mean}$, which implies that the range of magnetic fields
can extend beyond the mean free path of the quark.  Therefore,
successive exchanges of transverse gluons may correspond to
scatterings off the same parton. This interpretation is supported by
the infrared study performed in \cite{AurenGZ1}, where we found that
only transverse gluons are causing trouble in the infrared sector, and
that this problem occurs when $\lambda_{\rm coh} > \mu^{-1}$ (see
Eq.~(\ref{eq:condition-on-mu})). This condition means that the
production of a single photon occurs on a distance larger than the
correlation length of magnetic fields: this emission process is
therefore able to probe the scale of the magnetic screening, and the
rate is expected to become non perturbative. This qualitative
difference of transverse gluon exchanges is also supported by the fact
that the contribution of longitudinal gluons to $\Gamma$ is
perturbative (saturated at 1-loop), while the transverse contribution
to $\Gamma$ is non-perturbative \cite{BlaizI1,BlaizI2}.

\subsection{Comparison with the approach of Cleymans et al.}
The LPM effect has already been studied in the context of photon
production by a quark-gluon plasma, although with a very different
approach, in \cite{CleymGR2}. In this paper, the study follows the
semi-classical approach of \cite{Migda1}, to first deduce the photon
rate from a single quark trajectory, and then average over the
possible trajectories. Finally, the authors of \cite{CleymGR2} manage
to rewrite the rate as a function of a quantity
$F({\mbox{\boldmath$\theta$}},\tau)$ which is related to the
probability for the quark trajectory to undergo an angular deviation
$\mbox{\boldmath$\theta$}$ after a time $\tau$. Additionally, they
show that this object satisfies the following Fokker-Planck equation,
which would be in our notations\footnote{In \cite{CleymGR2}, only soft
  photons are considered. As a consequence, in this comparison, we
  assume always $q_0\ll p$, and therefore do not distinguish $p$ and
  $p+q_0$.}:
\begin{equation}
{{\partial F}\over{\partial \tau}}
+i{{q_0}\over{2}}\left[{\mbox{\boldmath$\theta$}}^2+
{{M^2_\infty}\over{p^2}}+{{Q^2}\over{q_0^2}}
\right]F=
{{\left<\mbox{\boldmath$\theta$}^2_{_{S}}\right>}\over 4}{\mbox{\boldmath$\nabla$}}_\theta^2 F\; ,
\end{equation}
where $\left<{\mbox{\boldmath$\theta$}}^2_{_{S}}\right>$ is the mean
square of scattering angle {\sl per unit length}.

In order to make the connection with our approach more transparent, we
can factorize out the quark width in the following way (recalling the
fact that $\Gamma^{-1}$ gives the mean free path):
\begin{equation}
\left<{\mbox{\boldmath$\theta$}}^2_{_{S}}\right>\equiv16\,\overline{{\mbox{\boldmath$\theta$}}^2} \Gamma\; ,
\end{equation}
where $\overline{{\mbox{\boldmath$\theta$}}^2}$ is the average square of the
scattering angle {\sl per collision} (up to a purely numerical factor)
and where the prefactor $16$ has been chosen for later convenience. We
can now rewrite the equation satisfied by $F$ in the following way
\begin{equation}
{{\partial F}\over{\partial \tau}}
+iq_0\left[
{{{\mbox{\boldmath$\theta$}}^2}\over 2}
+{1\over{2p^2}}\left\{
{\rm Re}\,M^2_{\rm eff}+i\,{\rm Im}\,M^2_{\rm eff}\,\overline{{\mbox{\boldmath$\theta$}}^2}
{\mbox{\boldmath$\nabla$}}_\theta^2
\right\}
\right]F=0\; .
\end{equation}
We see that this equation is governed by a complex number whose real
and imaginary part also appear in our approach. In particular, the
condition for having LPM effect (discussed after Eq.~(2.21) of
\cite{CleymGR2}) is the same as ours, given the fact that typical
scattering angles satisfy $\overline{{\mbox{\boldmath$\theta$}}^2}\sim
{\rm Re}\,M^2_{\rm eff}/p^2$.  Note however that our calculation also
includes the case of hard photon production, and shows that transverse
gluon exchanges are equally important, both points which are not
contained in Cleymans et al.'s approach.

\section{$q^*\bar{q}$ annihilation}
\label{sec:annihilation}
\subsection{Technical differences}
In section \ref{sec:2-loop-calcul}, when we performed the integral
over the quark energy $p_0$ using the $\delta(P^2-M^2_\infty)$ in
Eq.~\ref{eq:2-loop-start-simple}, we considered only the positive
energy solution, which corresponds to photon emission by the
bremsstrahlung of a quark. The calculation is very similar for the
contribution of $p_0=-\omega_{\imb p}$, with some peculiarities that
we are going to highlight in this section.

One of the most important differences is that, since we have $q_0>0$,
this contribution contains in fact the sum of two processes (see
Fig.~\ref{fig:processes}): the bremss\-trah\-lung of an antiquark if
$\omega_{\imb p}> q_0$, and the annihilation of an off-shell quark (a
quark put off-shell by a scattering) with an antiquark if
$\omega_{\imb p}<q_0$. The latter process has been studied in the case
of zero width and real photons in \cite{AurenGKZ1} and was indeed
found to be a very important contribution to hard photon production.
The purpose of this section is not to present a complete calculation
of this process (which would require the discussion of a few other
technical issues, especially in the region where $Q^2>4M^2_\infty$),
but only to reproduce in this case the discussion of section
\ref{sec:2-loop-discussion}, in order to have a qualitative picture of
the influence of the width on this process.

The calculation of the $p_0<0$ contribution can be performed by the
same method as the $p_0>0$ one. In particular, the angular integral
over $d\Omega_{\imb l}$ generates an effective mass $M^{\prime\, 2}_{\rm
  eff}$, the expression of which is
\begin{equation}
M^{\prime\, 2}_{\rm eff}\equiv M^2_\infty+{{Q^2}\over{q_0^2}}p(p-q_0)
+4i{\Gamma\over{q_0}}p(p-q_0)\; . 
\end{equation}
In addition to this change, the angular integral over $d\Omega_{\imb
  p}$ is now dominated by values of $\theta$ around $\pi$ (because
$1-\cos\theta$ in the expression of $\Delta$ becomes
$1+\cos\theta$)\footnote{This difference is also obvious when one
  compares the process on the left of Fig.~\ref{fig:processes} with
  the two processes on the right of the same figure, which we are
  dealing with now.}.  A practical consequence of this is that
$r\approx |p-q_0|$ in the collinear limit. All the subsequent steps
can be reproduced and one is lead to the following final formula:
\begin{eqnarray}
&&
{\rm Im}\Pi{}^\mu{}_\mu(Q)\approx {{2e^2 g^2}\over{\pi^4}}{T\over{q_0^2}}
\int\limits_{0}^{+\infty} dp\; {{p^2+(q_0-p)^2}\over 2}
[n_{_{F}}(q_0-p)-n_{_{F}}(-p)]
\nonumber\\
&&
\qquad\qquad\qquad\qquad\times
\sum\limits_{m=T,L}
\int\limits_{0}^{1}{{dx}\over x}\;\int\limits_{0}^{+\infty}dw\;
{{\left|\widetilde{I}_{m}\right|\,K(w,\widehat{\Gamma})}
\over
{(w+\widetilde{R}_{m})^2+(\widetilde{I}_{m})^2}}
\; ,
\label{eq:2-loop-final-neg-p0}
\end{eqnarray}
where the notations are the same as in
Eq.~(\ref{eq:2-loop-variables}), with $M^{\prime\, 2}_{\rm eff}$ in
place of $M^2_{\rm eff}$.

Another very important technical difference appears when one performs
the integral over the quark momentum $p$. Indeed, the range of this
integral is controlled by the statistical factors
$n_{_{F}}(r_0)-n_{_{F}}(p_0)\approx n_{_{F}}(p)-n_{_{F}}(p-q_0)$.
This factor is of order $1$ in a domain going from $p=0$ to $p\sim
{\rm Max}(q_0,T)$. In other words, if $q_0$ is soft, this range is the
usual $[0,T]$ interval. On the contrary, if $q_0\gg T$, the integral
extends to $p\sim q_0 \gg T$. A practical consequence is that for
$q_0\ll T$, the contribution of $p_0<0$ comes mainly from the
bremsstrahlung of an antiquark (since most of $dp$ integral satisfies
$q_0< p\sim T$), while for $q_o\gg T$ it is dominated by the
$q^*\bar{q}$ annihilation (since we then have $p< q_0$). In the
intermediate region $q_0\sim T$, this contribution is a mixture of
both processes.

\subsection{Modifications due to the width}
Again, we see that the term depending on $\Gamma$ in $M^{\prime\,
  2}_{\rm eff}$ comes with $\Gamma/q_0$, indicating that the width
will modify significantly the rate when $q_0$ is soft. But, in
addition, when $q_0\gg T$, $p$ can itself be of order $q_0$, so that
the imaginary part of $M^{\prime\, 2}_{\rm eff}$ is of order $\Gamma
q_0$, which can be much larger than $M^2_\infty$ at very large $q_0$.
It means that the width also changes dramatically the spectrum of
very hard real photons (in the case of bremsstrahlung, the width
modifies only marginally the rate of hard photons, since the imaginary
part of $M^{2}_{\rm eff}$ is of the same order of magnitude as
$M^2_\infty$ if $q_0\gg T$ and $\Gamma\sim g^2T$) coming from the
$q^*\bar{q}$ annihilation.

\begin{figure}[htbp]
\centerline{\rotatebox{-90}{\resizebox*{!}{7cm}{\includegraphics{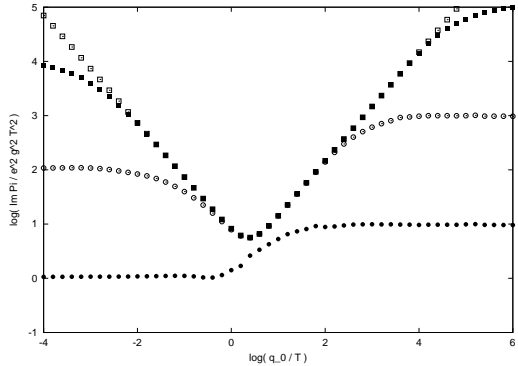}}}}
\caption{\sl Effect of the width in the region $p_0<0$ as a function 
  of $q_0/T$ (for $Q^2=0$). Each curve corresponds to a different
  value of the width $\Gamma$. The set of values taken for $\Gamma$ is
  the same as in Fig.~\ref{fig:width-effect}.}
\label{fig:width-effect-in-qq}
\end{figure}

This has been checked by evaluating numerically
Eq.~(\ref{eq:2-loop-final-neg-p0}), and the results are displayed in
Fig.~\ref{fig:width-effect-in-qq}, where we plot the imaginary part of
the photon polarization tensor as a function of $q_0/T$ (with the same
values of temperature and coupling constant as in
Fig.~\ref{fig:width-effect}) for a set of values of $\Gamma$ identical
to the one chosen in Fig.~\ref{fig:width-effect}.
From this plot, it is obvious that the regions $q_0\ll T$ and $q_0\gg
T$ are very different. In fact, in the region $q_0\ll T$, we obtain
results that exactly match those of Fig.~\ref{fig:width-effect}:
indeed, in this region, the $p_0<0$ contribution is dominated by the
bremsstrahlung of an antiquark, which as expected contributes equally
as the bremsstrahlung of a quark. 

The new feature coming with $p_0<0$ appears in the hard photon region
($q_0\gg T$), where now the $q^*\bar{q}$ annihilation contributes. At
very small $\Gamma$ (upper curve), we recover the result of
\cite{AurenGKZ1} according to which the process $q^*\bar{q}\to\gamma$
gives a contribution to ${\rm Im}\,\Pi{}^\mu{}_\mu(Q)$ that increases
like $q_0$. Then, as $\Gamma$ increases, we see a saturation at large
enough $q_0$, to a value that behaves like $\Gamma^{-1}$. This
confirms the above qualitative statement based on $M^{\prime\, 2}_{\rm
  eff}$. In this case also, it is possible to obtain a formula like
Eq.~(\ref{eq:2-loop-large-gamma}) showing explicitly the $\Gamma$
scaling law when $\Gamma$ is the dominant collinear regulator.

The contribution of the process $q^*\bar{q}\to \gamma$, first
evaluated at leading order in \cite{AurenGKZ1}, has been found to be
phenomenologically important for the production of direct photons in
the $2-10$ GeV range, because it dominates previous estimates by a
factor of order $5$ \cite{Sriva1}. The inclusion of this contribution
in heavy nuclei collisions simulations \cite{SrivaS1} has lead to a
good agreement with the measured rates from the WA98 experiment
\cite{Aggara1,Aggara2}. These rates did not include the LPM
suppression advocated here, but one must realize that the energy range
where thermal photons are relevant is also the region of minimum
sensitivity to the width $\Gamma$ (around the minimum of the curves in
figure \ref{fig:width-effect-in-qq}), since for a temperature of a
few hundred MeV, the GeV range corresponds to $q_0/T \sim 10$. It is
too early to be more quantitative here given the fact that vertex
corrections \cite{WorkIP1} have been completely disregarded in the
present work, but having a precise prediction for this process would
definitely be of important phenomenological interest.

\subsection{LPM effect in $q^*\bar{q}$ annihilation}
When $p_0<0$, the condition under which the width is the dominant
collinear regulator can be rewritten as
\begin{equation}
2\Gamma > {{q_0}\over{2}}
{{{\rm Re}\,M^{\prime\, 2}_{\rm eff}}\over{p|q_0-p|}}\; , 
\end{equation}
and the quantity in the right hand side is the minimal value $l_{\rm
  min}$ for the momentum $L$ of the gluon exchanged in the scattering.
Therefore, its inverse is also the coherence length for the emission
process, and the above condition is nothing but the criterion for
having LPM effect.

As a consequence, we are lead to the conclusion that the LPM effect
plays a role not only for the photon production by bremsstrahlung at
low energy, but also for the $q^*\bar{q}$ annihilation at high energy.
This conclusion was also recently obtained in \cite{BaierK2} by a
semi-classical method following the original method of Migdal
\cite{Migda1}. The authors of \cite{BaierK2} in fact studied the
inverse of the process $q^*\bar{q}\to\gamma$, namely the production of
a fermion pair out of a photon (this process is made possible by the
fact that one of the fermions is produced off-shell and then scatters
in the medium). They found that the LPM effect is important for the
fragmentation of very high energy photons. This is equivalent to our
observation that the LPM effect modifies $q^*\bar{q}\to\gamma$ for
large $q_0$.

\section{Conclusions}
In this paper, we have studied the effect of the finite mean free path of
quarks on photon production by a quark gluon plasma. The main result
is that such a correction may have important effects by affecting the
collinear sector.

We have first considered the effect of a quark width in the simplest
1-loop diagram, and shown that potentially large 1-loop contributions
cancel when a more complete summation is performed. This cancellation
is not new in fact, and has been noticed elsewhere. It has
connections with the fact that there are no hard thermal loops for
vertices with 2 photons and any number of gluons.

Two-loop contributions beyond the eikonal approximation escape this
cancellation, and lead to a result which exhibits features of the LPM
effect. It is indeed possible to interpret the region where the width
of the quark is important as the region where the formation length of
the photon is larger than the mean free path of the quark. Our study also
shows that the LPM effect modifies the emission spectrum of very hard
photons.  Despite its nice interpretation, our present result is not
complete because it does not consider the vertex corrections that
should accompany the resummation of a width.

Multiple scatterings due to longitudinal gluon exchanges are
independent, as in the semi-classical treatment of the LPM effect.
On the contrary, the transverse sector displays features that are
qualitatively different from what is usually accepted in the
semi-classical treatment (which just disregards transverse gluon
exchanges). Indeed, we find that multiple transverse gluon exchanges
are very important also, and that they are not independent due to the
long range of magnetic fields.

\begin{figure}[htbp]
\centerline{\resizebox*{!}{4cm}{\includegraphics{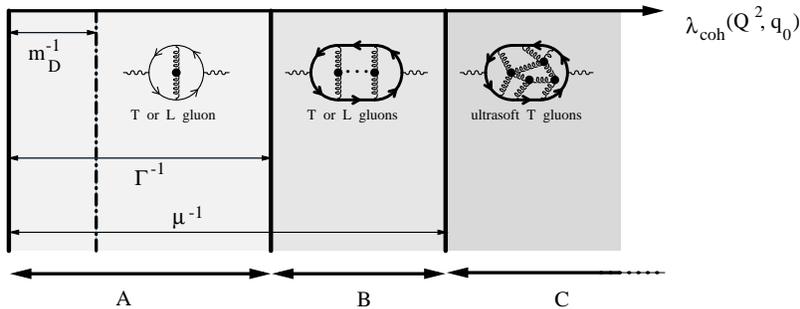}}}
\caption{\sl Summary of the various scales in the problem, and 
  comparison with the coherence length.  In region A: $0<\lambda_{\rm
    coh}<\lambda_{\rm mean}$.  Region B: $\lambda_{\rm mean}<
  \lambda_{\rm coh}< \mu^{-1}$. Region C: $\mu^{-1}< \lambda_{\rm
    coh}$. Each picture shows the new class of diagrams that one must
  consider when going to larger and larger coherence lengths (in
  addition to the diagrams already considered for smaller coherence
  lengths). A boldface quark line denotes the resummation of a width
  in addition to the asymptotic thermal mass.}
\label{fig:scales}
\end{figure}

The various scales in the problem are summarized in
Fig.~\ref{fig:scales}. Three of these scales are intrinsic to the
quark gluon plasma: the Debye screening length $m_{_{D}}^{-1}$, the
mean free path of the quarks $\lambda_{\rm mean}\sim\Gamma^{-1}$ and
the magnetic screening length $\mu^{-1}$.  The fourth scale, the
coherence length $\lambda_{\rm coh}$ (or formation length of the
photon), depends on the energy and invariant mass of the photon one
wants to observe, and should be compared to the first three.
This leads to three zones which have simple physical interpretations.
In addition, the nature (and complexity) of the dominant higher loop
corrections depend on the coherence length.

\noindent $\bullet$ Region A: $0<\lambda_{\rm coh}<\lambda_{\rm mean}$, and 
photon production is dominated by single scatterings. The large scale
structure of magnetic fields is irrelevant. The only relevant diagram
is the 2-loop diagram (with HTL resummed quarks) already considered in
\cite{AurenGKP2}. The contributions of transverse and longitudinal
gluons are comparable.

\noindent $\bullet$ Region B: $\lambda_{\rm mean}< 
\lambda_{\rm coh}< \mu^{-1}$, which implies that photon production is
affected by multiple scatterings (LPM effect). In this region, the
emission process is not yet sensitive to the magnetic screening.
Indeed, according to \cite{AurenGZ1}, this sensitivity comes in when
$\lambda_{\rm coh} > \mu^{-1}$.  It is sufficient to resum a width
(saturated at 1-loop\footnote{When the IR cutoff is $\lambda_{\rm
    coh}^{-1}\gg\mu$, corrections to $\Gamma$ coming from topologies
  beyond 1-loop are suppressed by $g^2T/\lambda_{\rm coh}^{-1}\ll
  1$.}) on the quarks, and to consider vertex corrections involving
both longitudinal and transverse gluons. So far, we have said nothing
about the vertex corrections that should come with self-energy
insertions. Nevertheless, it is reasonable to assume from Ward
identities that 1-loop self-energy insertions talk to ladder
corrections\footnote{The mechanism which makes these ladder
  corrections important is the same as in \cite{LebedS1,LebedS2}, and
  has nothing to do with infrared singularities. The arguments of
  \cite{AurenGZ1} cannot exclude them, even if the infrared cutoff is
  $\lambda_{\rm coh}^{-1}\gg\mu$. This is why these corrections are
  due to both longitudinal and transverse gluons.}. We can note also
that the discussion of this region is somewhat academic, because its
extension is only proportional to $\ln(1/g)$. Additionally, the new
topologies of the region C are at most suppressed by powers of
$\ln(1/g)^{-1}$ if evaluated in region B.

\noindent $\bullet$ Region C: $\mu^{-1}< \lambda_{\rm coh}$. The LPM 
effect still modifies photon production. In addition, the emission of
a photon lasts long enough for the process to be sensitive to the
magnetic screening. This is the physical meaning of the result of
\cite{AurenGZ1}. Diagrams with an arbitrary number of ultrasoft
transverse gluons, connected in all the possible ways, must be
resummed. In addition, the width included on the quarks contains non
perturbative contributions due to transverse gluons
\cite{BlaizI1,BlaizI2}, and the effective gluon vertices must be
corrected to hold for ultrasoft momenta \cite{BlaizI9}.

A full study (including vertex corrections) of the corrections to
photon production due to longitudinal gluons seems within the reach of
perturbation theory, but the situation is very different in the
transverse sector which seems far beyond the possibilities of
perturbative methods. In this respect, photon production is not very
different from the calculation of the quark damping rate
\cite{BlaizI1,BlaizI2}. New tools, like functional methods, transport
equations, and eventually lattice techiques, are presumably the way to
go in this area. In particular, the picture emerging from
Fig.~\ref{fig:scales} suggests an analogy with the successive
resummations of \cite{Bodek2,Bodek3}, in which the scales $T$, $gT$
and $g^2T\ln(1/g)$ are successively integrated out. Indeed, going to
larger and larger coherence lengths requires the inclusion of more and
more complicated topologies, a procedure which amounts to integrate
out degrees of freedom encountered at smaller length scales.

\section*{Acknowledgements}
F.G. would like to thank R. Pisarski, A. Peshier, R. Venugopalan, D.
B\"odeker and S. Jeon for very interesting comments and discussions.
H.Z. thanks the Institute of Nuclear Theory at the University of
Washington for its hospitality where part of this work has been done
and Dam Thanh Son for fruitful discussions. The work of F.G. is
supported by DOE under grant DE-AC02-98CH10886.

\appendix

\section{Separation of the various cuts}
\label{app:cuts}
Equation (\ref{eq:2-loop-interm}) includes automatically the sum of
the cuts $(c)$ and $(d)$ of Fig. \ref{fig:diagrams}. To see it
explicitly, let us first denote
\begin{eqnarray}
&&F(\Gamma)\equiv {1\over{u+{{M^2_{\rm eff}}\over{2p^2}}}}\nonumber\\
&&G(\Gamma)\equiv{{\epsilon(\Gamma)}\over{\Big[
\Big(
u+{{M^2_{\rm eff}}\over{2p^2}}+{{L^2}\over{2p^2}}
\Big)^2-{{L^2}\over{p^2}}{{M^2_{\rm eff}}\over{p^2}}
\Big]^{1/2}}}\; .
\end{eqnarray}
Then, the discontinuity on the last line of
Eq.~(\ref{eq:2-loop-interm}) equals:
\begin{eqnarray}
&&{\rm Disc}\,F(\Gamma)G(\Gamma)
=F(\Gamma)G(\Gamma)-F(-\Gamma)G(-\Gamma)\nonumber\\
&&\qquad={{[G(\Gamma)+G(-\Gamma)]}\over{2}}\;{\rm Disc}\,F(\Gamma)
+{{[F(\Gamma)+F(-\Gamma)]}\over{2}}\;{\rm Disc}\,G(\Gamma)\; .
\end{eqnarray}
In the last line, the first term corresponds to cut $(d)$ and the
second term gives cut $(c)$. The fact that both cuts contribute to the
photon rate when $\Gamma\not=0$ was to be expected. Indeed, the cut
$(d)$ which corresponds to the process $q\to q\gamma$ is kinematically
forbidden if $\Gamma=0$, but is now allowed by the very fact that the
width $\Gamma$ reflects the collisions of the quasi-quark (and the
channel $q\to q\gamma$ is allowed in the presence of a medium).

The integrals to be calculated are more complicated if one wants the
two cuts separately. But it turns out that the difference between the
two cuts is also very simple. This difference is given by:
\begin{eqnarray}
&&(c)-(d)\propto{{[F(\Gamma)+F(-\Gamma)]}\over{2}}\;{\rm Disc}\,G(\Gamma)-
{{[G(\Gamma)+G(-\Gamma)]}\over{2}}\;{\rm Disc}\,F(\Gamma)
\nonumber\\
&&\qquad=F(-\Gamma)G(\Gamma)-F(\Gamma)G(-\Gamma)
\; .
\end{eqnarray}
Therefore, in order to get $(c)-(d)$, one has to start from
Eq.~(\ref{eq:2-loop-interm}) in which the last line is substituted by
${\rm Disc}\,F(-\Gamma)G(\Gamma)$. The same transformations can be
applied to this integral, and one finally obtains an equation similar
to Eq.~(\ref{eq:2-loop-final}), but in which the function
$K(w,\widehat{\Gamma})$ is replaced by
\begin{equation}
L(w,\widehat{\Gamma})\equiv
{1\over 2}\int\limits_{0}^{1}{{dy}\over{\sqrt{1-y}}}
{{y+4/w}\over{(y+4/w)^2+(1-y)(4\widehat{\Gamma}/w)^2}}\; .
\end{equation}
This new integral is also elementary and can be obtained in closed
form if needed.

It is obvious that in the limit of zero width $(\widehat{\Gamma}\to
0)$, the two integrals $K(w,\widehat{\Gamma})$ and
$L(w,\widehat{\Gamma})$ become equal. The only way for this to happen
is that the contribution of cut $(d)$ goes to zero.  Therefore, we
recover the fact that the process $q\to q\gamma$ disappears when
$\Gamma=0$.



\end{document}